# RbFe$^{2+}$Fe$^{3+}$F$_6$: Synthesis, Structure, and Characterization of a New Charge-Ordered Magnetically Frustrated Pyrochlore-Related Mixed-Metal Fluoride

Sun Woo Kim,[a] Sang-Hwan Kim,[a] P. Shiv Halasyamani,[a,*] Mark A. Green,[b] Kanwal Preet Bhatti,[c] C. Leighton,[c,*] Hena Das,[d] and Craig J. Fennie[d,*]

A new charge-ordered magnetically frustrated mixed-metal fluoride with a pyrochlore-related structure has been synthesized and characterized. The material, RbFe$_2$F$_6$ (RbFe$^{2+}$Fe$^{3+}$F$_6$) was synthesized through mild hydrothermal conditions. The material exhibits a three-dimensional pyrochlore-related structure consisting of corner-shared Fe$^{2+}$F$_6$ and Fe$^{3+}$F$_6$ octahedra. In addition to single crystal diffraction data, neutron powder diffraction and magnetometry measurements were carried out. Magnetic data clearly reveal strong antiferromagnetic interactions (a Curie-Weiss temperature of -270 K) but sufficient frustration to prevent ordering until 16 K. No structural phase transformation is detected from the variable temperature neutron diffraction data. Infrared, UV-vis, thermogravimetric, and differential thermal analysis measurements were also performed. First-principles density functional theory (DFT) electronic structure calculations were also done. Crystal data: RbFe$_2$F$_6$, orthorhombic, space group Pnma (No. 62), a = 7.0177(6) Å, b = 7.4499(6) Å, c = 10.1765(8) Å, V = 532.04(8) Å$^3$, Z = 4.

## Introduction.

Mixed-metal fluorides are of topical interest attributable to their varied functional properties.[1] These include multiferroic behavior (BaNiF$_4$[2-5] and Pb$_5$Cr$_3$F$_{19}$[6]), magnetic frustration, (Na$_2$NiFeF$_7$[7] and MnCrF$_5$[8]), ferroelectricity, (K$_3$Fe$_5$F$_{15}$[9-10] and SrAlF$_5$[11]), and non-linear optical behavior (BaMgF$_4$).[12-13] Recently, attention has been paid to the multiferroic K$_3$Fe$_5$F$_{15}$ and related materials such as K$_3$Cu$_3$Fe$_2$F$_{15}$ and K$_3$Cr$_2$Fe$_2$F$_{15}$,[14-16] as well as multiferroic fluorides as a whole.[17-18] K$_3$Fe$_5$F$_{15}$ and K$_3$Cu$_3$Fe$_2$F$_{15}$ have been shown to be ferri- and anti-ferromagnetic respectively, whereas K$_3$Cr$_2$Fe$_2$F$_{15}$ exhibits relaxor-like magnetic transitions. Although full structural data is lacking for the quaternary phases, multiferroic behavior is suggested.[15-16] It should also be noted that in K$_3$Cu$_3$Fe$_2$F$_{15}$ and K$_3$Cr$_2$Fe$_2$F$_{15}$, the Cu$^{2+}$ / Fe$^{3+}$ and Cr$^{3+}$ / Fe$^{2+}$ cations respectively are crystallographically disordered.

With respect to AM$^{2+}$M$^{3+}$F$_6$ materials (A = alkali metal or NH$_4$; M$^{2+}$ = Mg, Mn, Fe, Co, Ni, Cu; M$^{3+}$ = Al, Ga, V, Cr, Fe) a host of materials have been reported,[19-20] although well-determined crystal structures are lacking for many. Structure types for the AM$^{2+}$M$^{3+}$F$_6$ materials include trirutile (LiM$^{2+}$M$^{3+}$F$_6$),[21] modified pyrochlore ((NH$_4$)Fe$^{2+}$Fe$^{3+}$F$_6$),[22] tetragonal (K$_{0.6}$Fe$^{2+}_{0.6}$Fe$^{3+}_{0.4}$F$_3$),[23-25] and hexagonal (K$_{0.6}$Nb$_2$F$_6$)[26] bronzes, and materials iso-structural to trigonal Na$_2$SiF$_6$ (LiMnGaF$_6$).[27] In these materials, both disorder and order are observed between the M$^{2+}$ and M$^{3+}$ cations. With the pyrochlore related materials, crystallographic disorder, of the M$^{2+}$ and M$^{3+}$ cations on the octahedral sites is observed. This disorder results in spin-glass behavior in CsMnFeF$_6$.[28-29] Ordering of the M$^{2+}$ and M$^{3+}$ cations has been observed, with a lowering of crystallographic symmetry, in the trirutile LiFe$^{2+}$Fe$^{3+}$F$_6$,[21] the fluorobronze K$_{0.6}$Fe$^{2+}_{0.6}$Fe$^{3+}_{0.4}$F$_3$,[23-25] and the pyrochlore-related (NH$_4$)Fe$^{2+}$Fe$^{3+}$F$_6$.[22] Antiferromagnetic behavior has been observed with the Li$^+$ and NH$_4$ phases.[21, 30] In addition to the aforementioned magnetic behavior, magnetic frustration has been observed in a variety of mixed-metal fluorides.[1] Such frustration can occur not only when the two metal cationic species, M$^{2+}$ and M$^{3+}$, crystallographically order, but also if they are arranged in some form of triangular structural topology, i.e. in the presence of geometric frustration. Magnetically frustrated fluorides include the hexagonal tungsten bronze – FeF$_3$,[31] Na$_2$NiFeF$_7$,[7] MnCrF$_5$,[8] Fe$_3$F$_8$ · 2H$_2$O,[32-33] and NH$_4$Fe$_2$F$_6$.[30]

In this paper, we report on the synthesis, structure (X-ray and variable temperature neutron diffraction), and characterization of RbFe$_2$F$_6$ (RbFe$^{2+}$Fe$^{3+}$F$_6$). This new material represents an example of a charge-ordered pyrochlore-related mixed-metal fluoride that exhibits strong magnetic frustration. In addition to the synthesis and structural characterization, magnetic measurements and theoretical calculations are performed. These measurements and calculations enable us to develop and understand a variety of important structure-property relationships.

## Experimental Section

**Reagents.** RbF (Alfa Aesar, 99.7%), FeF$_2$ (Alfa Aesar, 99%), FeF$_3$ (Alfa Aesar, 97%), and CF$_3$COOH (Alfa Aesar, 99%) were used without further purification.

**Synthesis.** RbFe$_2$F$_6$ was obtained by hydrothermal methods using a diluted CF$_3$COOH solution. 0.119 g (1.14×10$^{-3}$ mol) of RbF, 0.107 g (1.14×10$^{-3}$ mol) of FeF$_2$, 0.129 g (1.14×10$^{-3}$ mol) of FeF$_3$, 3 ml (3.90×10$^{-2}$ mol) of CF$_3$COOH, and 5 ml of H$_2$O were combined in a 23-mL Teflon-lined stainless steel autoclave. The autoclave was closed, gradually heated to 230 ºC, held for 24 h, and cooled slowly to room temperature at a rate 6 ºC h$^{-1}$. The mother liquor was decanted, and the only solid product from the reaction, brown colored rod shaped crystals, subsequently shown to be RbFe$_2$F$_6$, was recovered by filtration and washed with distilled water and acetone. The yield was ~40 % based on FeF$_3$. Powder X-ray diffraction patterns on the synthesized phase are in good agreement with the generated pattern from the single-crystal data (see Figure S1).

**Single Crystal X-ray Diffraction.** A brown colored rod shaped crystal (0.02 x 0.02 x 0.1 mm$^3$) was selected for single-crystal data collection. The data were collected using a Siemens SMART APEX diffractometer equipped with 1K CCD area detector using graphite-monochromated Mo-Kα radiation. A hemisphere of data was collected using a narrow-frame method with scan widths of 0.30° in ω and an exposure time of 45 s per frame. The data were integrated using the Siemens SAINT program,[34] with the intensities corrected for Lorentz, polarization, air absorption, and absorption attributable to the variation in the path length through the detector face plate. Psi-scans were used for the absorption correction on the data. The data were solved and refined using SHELXS-97 and SHELXL-97,[35-36] respectively. All of the atoms were refined with anisotropic thermal parameters and the refinement converged for I > 2σ(I). All calculations were performed using the WinGX-98 crystallographic software package.[37] Relevant crystallographic data, atomic coordinates and thermal parameters, and selected bond distance for $RbFe_2F_6$ are given in Tables 1, 2, and 3.

**Table 1.** Crystallographic Data for $RbFe_2F_6$

| Parameter | $RbFe_2F_6$ |
|---|---|
| Formula Weight, fw | 311.17 |
| T (K) | 296.0(2) |
| λ (Å) | 0.71073 |
| Crystal System | Orthorhombic |
| Space Group | Pnma (No.62) |
| a (Å) | 7.0177(6) |
| b (Å) | 7.4499(6) |
| c (Å) | 10.1765(8) |
| V (Å$^3$) | 532.04(8) |
| Z | 4 |
| $\rho_{calcd}$ (g/cm$^3$) | 3.885 |
| $\mu$(mm$^{-1}$) | 14.577 |
| $2\theta_{max}$ (deg) | 58.04 |
| R (int) | 0.0361 |
| GOF | 1.087 |
| R (F)$^a$ | 0.0214 |
| $R_w(F_o^2)^b$ | 0.0497 |

$^a R(F) = \Sigma ||F_o| - |F_c|| / \Sigma |F_o|$, $^b R_w(F_o^2) = [\Sigma w(F_o^2 - F_c^2)^2 / \Sigma w(F_o^2)^2]^{1/2}$

**Table 2.** Atomic Coordinates for $RbFe_2F_6$

| Atom | x | Y |
|---|---|---|
| Rb(1) | 0.9920(1) | 0.25 |
| Fe(1) ($Fe^{2+}$) | 0.7967(1) | 0.25 |
| Fe(2) ($Fe^{3+}$) | 0.5 | 0 |
| F(1) | 0.7364(2) | 0.0633(2) |
| F(2) | 0.3737(2) | 0.0104(2) |
| F(3) | 0.4359(3) | 0.25 |
| F(4) | 0.5642(3) | 0.25 |

$^a$ $U_{eq}$ is defined as one-third of the trace of the orthogonal $U_{ij}$ tensor.

**Table 3.** Selected Bond Distances for $RbFe_2F_6$

| Bond | Distance (Å) | Bond |
|---|---|---|
| Rb(1) — F(1) | 3.042(1) × 2 | Fe(1) — F(1) |
| Rb(1) — F(1) | 3.071(1) × 2 | Fe(1) — F(2) |
| Rb(1) — F(2) | 3.050(2) × 2 | Fe(1) — F(4) |
| Rb(1) — F(2) | 3.248(1) × 2 | Fe(1) — F(4) |
| Rb(1) — F(3) | 3.239(2) | Fe(2) — F(1) |
| Rb(1) — F(4) | 2.931(2) | Fe(2) — F(2) |
|  |  | Fe(2) — F(3) |

**Powder X-ray Diffraction.** The PXRD data of $RbFe_2F_6$ were collected on a PANalytical X'pert pro diffractometer using Cu Kα radiation in the 2θ range 5-90°. A step size of 0.008 degrees (deg) with a scan time of 0.3 s/deg was used. No impurity phases were observed, and the calculated and experimental PXRD patterns are in good agreement (see Figure S1).

**Neutron Diffraction.** Powder neutron diffraction was performed on the BT1 high resolution diffractometer at the NIST Center for Neutron research. Data were collected using a Ge (311) monochromator at λ = 2.0782 Å and a (311) monochromator at λ = 1.5401 Å, with an in-pile collimation of 15'. Rietveld refinements were performed using the FULLPROF suite of programs.[38] Cooling was performed with a closed cycle refrigerator and measurements were performed at 4, 10, 25, 50, 100, 150, 200, 250 and 300 K.

**Infrared Spectroscopy.** Infrared spectra were collected on a Mattson FT-IR 5000 spectrometer in the 400 – 4000 cm$^{-1}$ range (see Figure S3).

**UV-Vis Diffuse Reflectance Spectroscopy.** UV-Vis diffuse reflectance spectra were collected on a Varian Cary 500 UV-Vis-NIR spectrophotometer from 200-1500 nm at room temperature. Poly-(tetrafluoroethylene) was used as a reference. Reflectance spectra were converted to absorbance using the Kubelka-Munk function (see Figure S4).[39-40]

**Thermal Analysis.** Thermogravimetric analysis was carried out on a EXSTAR TG/DTA 6300 (SII NanoTechnology Inc.). About 10 mg of the sample was placed into a platinum crucible and heated under a nitrogen atmosphere at a rate of 10 °C min$^{-1}$ to 1000 °C (see Figure S5).

**Magnetic Measurements.** DC magnetometry measurements were performed in helium gas in a commercial SQUID (Superconducting Quantum Interference Device) magnetometer (Quantum Design) at temperatures from 4.0 to 300 K, in applied magnetic fields up to 70 kOe. For low field measurements the remnant field in the superconducting magnet was nulled using a compensating coil in unison with a fluxgate.

**Theoretical Calculations.** First-principles electronic structure calculations based on density functional theory (DFT) were performed using the Perdew-Burke-Ernzerhof (PBE)[41] form of exchange correlation functional under the generalized gradient approximation (GGA). The projected augmented plane wave (PAW)[42] method was employed as implemented in the Vienna Ab-initio Simulation Package (VASP).[43-44] The valence electron configurations used in the present study are follows: $4p^6 5s^1$ for Rb, $3p^6 3d^7 4s^1$ for Fe and $2s^2 2p^5$ for F. We used a 500 eV plane wave cutoff to generate the basis set, and a 4×4×2 Monkhorst-Pack k-point mesh to sample the electronic Brillouin zone. The optimization of the internal structural parameters were carried out and optimized until the forces on each atom of the system become < 0.001 eV/Å. Within density-functional theory, the failure of the generalized gradient approximation (GGA) to properly capture the physics of correlated systems is well established. A widely accepted approach beyond GGA is the GGA plus Hubbard U (GGA+U) method.[45] We employed the rotationally invariant scheme proposed by Dudarev et al.[46] in which a single parameter is $U_{eff}$ = U - JH, where U and JH represent the effective on-site screened Coulomb and exchange interaction is introduced on the Fe-atom. The structural, electronic and magnetic properties of RbFe$_2$F$_6$ have been investigated both in GGA and in GGA+U for the range of $U_{eff}$ equal to 1.0 to 7.0 eV. Qualitatively, all results remain insensitive to $U_{eff}$ > 4.0eV therefore we present results for GGA and GGA+U = 6eV (See supplemental for entire complete study, from it a high level of confidence is obtained that a suitable value of the on-site Coulomb interaction lies between 6.0 to 7.0 eV. Such fine-tuning of U does not alter the physics discussed here.)

## Results

**Structure.** RbFe$_2$F$_6$ crystallizes in the Pnma space group with lattice parameters of a = 7.02134(7) Å, b = 7.45093(7) Å and c = 10.1795(1) Å at room temperature. This represents a reduced orthorhombic cell of the conventional β-pyrochlore lattice with the Fd-3m symmetry. The β-pyrochlore is related to the more common α-pyrochlore lattice, with the general formula A$_2$B$_2$X$_6$X', through two ordered vacancies. Firstly, one of the two A cations is vacant, which reverts the Kagome network of the A sites in α-pyrochlore into a diamond lattice with T$_d$ point symmetry. Secondly, the anion, X', that does not contribute to the BX$_6$ octahedra is vacant; removing this apical anion reduces the eight-coordination of the A cation and leaves an open cage site.

RbFe$_2$F$_6$ exhibits a three-dimensional crystal structure consisting of corner-shared FeF$_6$ octahedra that are separated by Rb$^+$ cations (see Figures 1 and 2). The formula may be more descriptively written as RbFe$^{2+}$Fe$^{3+}$F$_6$, as the Fe$^{2+}$ and Fe$^{3+}$ cations are ordered in the structure. The structure of RbFe$_2$F$_6$ may be described as being built up from two connected FeF$_6$ octahedral sub-lattices. Polyhedral and ball-and-stick representations of RbFe$_2$F$_6$ are shown in Figures 1 and 2. The bc-plane of the structure is shown in Figure 1, and as can be seen chains of Fe$^{3+}$F$_6$ octahedra share corners along the b-axis direction. These Fe$^{3+}$F$_6$ chains are connected through Fe$^{2+}$F$_6$ octahedra along the c-axis direction. The ac-plane of the structure is shown in Figure 2. Similarly,

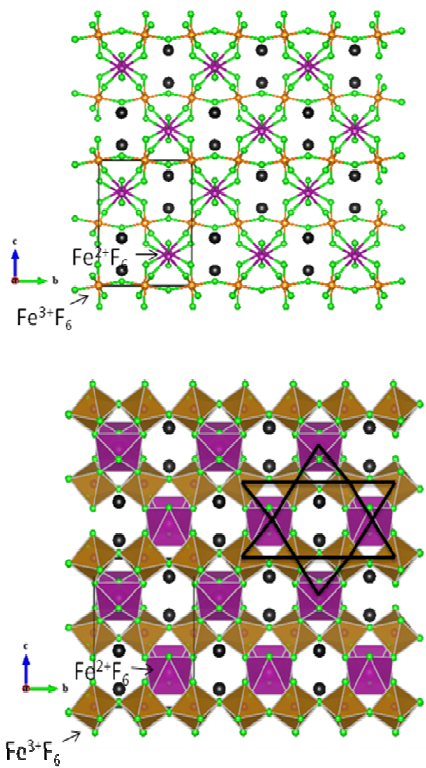

**Figure 1.** Ball-and-stick (top) and polyhedral (bottom) representations of RbFe$_2$F$_6$ in the bc-plane. Note the Kagome-type nets in the bottom figure (dark lines).

chains of Fe$^{2+}$F$_6$ octahedra share corners along the *a*-axis direction, and these chains are connected through Fe$^{3+}$F$_6$ octahedra along the *c*-axis direction. This octahedral connectivity results in Kagome-type nets in *both* the bc- and ac-planes of the structure (see Figures 1b and 2b). The Fe$^{2+}$ - F (Fe$^{3+}$ - F) bond distances range between 1.961(2)-2.1368(13)Å (1.9098(14)-1.9488(6)Å). The Rb$^+$ cation is in a 10-fold coordinated environment, with Rb-F distances that range between 2.931(2)-3.2477(14) Å. In connectivity terms, the structure may be written as {(Fe(II)F$_{6/2}$)$^-$ (Fe(III)F$_{6/2}$)$^0$}$^-$ where charge balance is maintained by a Rb$^+$ cation. Bond valence calculations[47-48] (see Table 4) result in values of 0.821, 1.93, 3.03 and 0.935-0.996 for

$Rb^+$, $Fe^{2+}$, $Fe^{3+}$ and $F^-$ respectively.

Table 4. Bond valence analysis for $RbFe_2F_6$[a]

| Atom | F(1) | F(2) | F(3) | F(4) | $\sum_{cations}$ |
|---|---|---|---|---|---|
| **Rb(1)** | $0.092^{[\times 2]}$ $0.085^{[\times 2]}$ | $0.090^{[\times 2]}$ $0.053^{[\times 2]}$ | 0.054 | 0.124 | **0.818** |
| **Fe(1)** | $0.288^{[\times 2]}$ | $0.268^{[\times 2]}$ | - | 0.431 0.380 | **1.92** |
| **Fe(2)** | $0.533^{[\times 2]}$ | $0.536^{[\times 2]}$ | $^{[\times 2]}0.482^{[\times 2]}$ | - | **3.10** |
| $\sum_{anions}$ | **0.998** | **0.947** | **1.02** | **0.935** | |

[a] Bond Valence sums calculated with the formula: $S_i = \exp[(R_0 - R_i)/B]$, where $S_i$ is a valence of the bond "$i$", $R_0$ is a constant dependent upon the bonded elements, $R_i$ is the bond length of bond $i$ and $B$ equals 0.37. Left and right superscripts indicate the number of equivalent bonds for anions and cations, respectively.

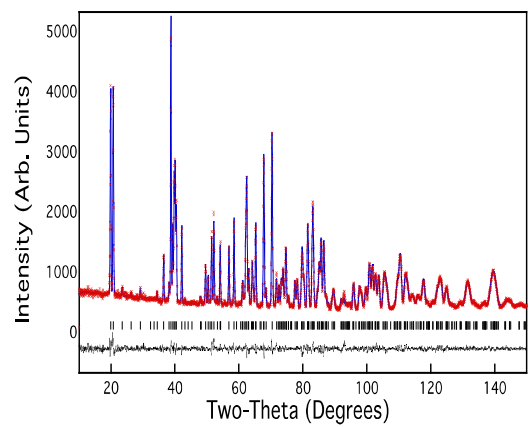

**Figure 3.** Observed (red), calculated (blue) and difference (black) data obtained from Rietveld refinements of neutron diffraction data of $RbFe_2F_6$ at 300 K.

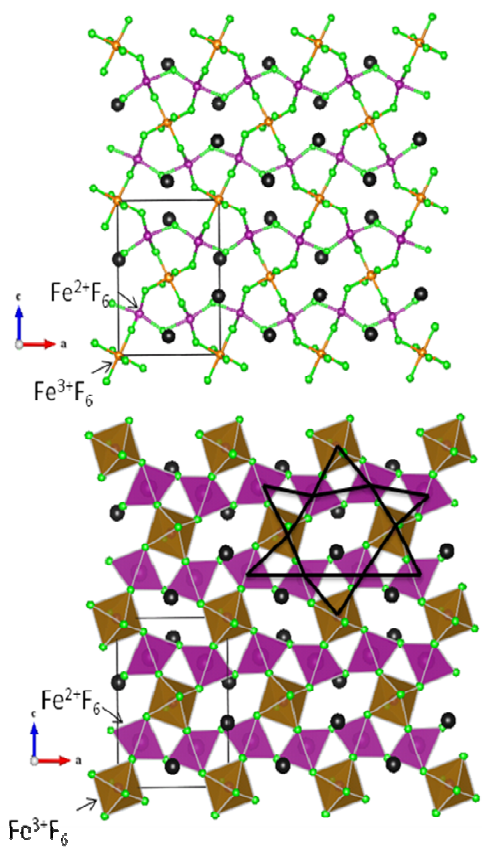

**Figure 2.** Ball-and-stick (top) and polyhedral (bottom) representations of $RbFe_2F_6$ in the ac-plane. Note the Kagome-type nets in the bottom figure (dark lines).

**Neutron Diffraction.** Powder neutron diffraction measurements were performed on $RbFe_2F_6$ at several temperatures in the range 4 - 300 K. Figure 3 shows the typical quality of fit to the observed data with this model, as obtained at 300 K, resulting in goodness-of-fit factors of wRp = 4.49 and $\chi^2$ = 1.29. One key feature of β-pyrochlores that has greatly hindered their usefulness as model magnetic systems is their tendency for both site disorder and partial occupancy. To evaluate these possibilities a number of models were tested and no evidence could be found for such issues in $RbFe_2F_6$; varying the occupancy of Rb from the ideal value of 1.0 gave a refined value of 0.99(1) and, as this made no improvement to the goodness-of-fit factors, this parameter was fixed in all subsequent refinements.

**Infrared Spectroscopy.** The FT-IR spectra of $RbFe_2F_6$ revealed Fe-F vibrations between 1000 and 400 $cm^{-1}$. The bands occurring between 750 – 700 $cm^{-1}$ and 530 – 400 $cm^{-1}$ can be assigned to Fe – F and Fe – F – Fe vibrations, respectively. These assignments are consistent with previous reports.[49] The IR spectra and assignments have been deposited in the supporting information (see Figure S3).

**UV-Vis Diffuse Reflectance Spectroscopy.** The UV-vis diffuse reflectance spectra indicate that the absorption energy for $RbFe_2F_6$ is approximately 1.9eV, consistent with the brown color of the material. Absorption (K/S) data were calculated through the Kubelka-Munk function:

$$F(R) = (1 – R)^2 / 2R = K/S$$

where $R$ represents the reflectance, $K$ the absorption, and $S$ the scattering. In a $K/S$ versus E(eV) plot, extrapolating the linear part of rising curve to zero provides the onset of absorption at 1.9 eV. One of three bands in the region of 1.5 – 3.3 eV were attributed to d-d transitions of Fe, the other large broad bands in the region of 3.5 – 5.0 eV were attributed to metal to ligand charge transfer. The UV-vis diffuse reflectance spectra have been deposited in the supporting information (see Figure S4).

**Thermal Analysis.** The thermal behavior of $RbFe_2F_6$ was investigated using thermogravimetric analysis (TGA) and differential thermal analysis (DTA) under a $N_2$ atmosphere. The decomposition started around 350 °C and an additional step was also observed at around 600 °C, which is likely attributable to the loss of fluorides. The DTA also showed two endothermic peaks at ~450 °C and ~750 °C, that indicate decomposition. Thermogravimetric analysis and differential thermal analysis diagrams for RbFe$_2$F$_6$ have been deposited in the supporting information, Figure S5. The final residue products, RbFeF$_3$ and FeF$_2$, were confirmed by PXRD (see Figure S6).

**Magnetic Measurements.** Figure 4 shows a summary of the basic magnetic characterization of a powder sample of RbFe$_2$F$_6$. The data shown are measured in an applied magnetic field ($H$) of 1 kOe, after field cooling (FC) and zero field cooling (ZFC). As shown in panel (a) the d.c. magnetic susceptibility ($\chi$) is positive, exhibiting a monotonic increase with decreasing temperature down to 16 ± 0.5 K, at which point a prominent peak occurs (see inset) and the FC and ZFC curves bifurcate. The $\chi(T)$ behavior is thus typical of an antiferromagnet, consistent with the low $T$ neutron diffraction analysis discussed below.

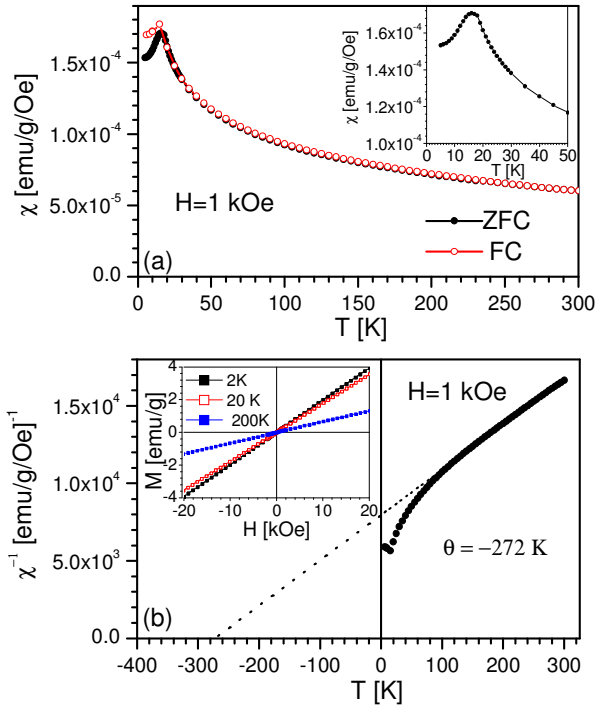

**Figure 4.** Temperature dependence of (a) the d.c. magnetic susceptibility measured in 1 kOe (after zero field and field cooling), and (b) the inverse magnetic susceptibility with a Curie-Weiss fit (dotted line). The extracted parameters are shown in the figure. The inset to (a) shows a close up of the low temperature region revealing the 16 K Néel temperature. The inset to (b) shows the linear magnetization vs. field behavior over the whole temperature range studied.

**Theoretical Calculations.** We optimized the internal structural degrees of freedom keeping the lattice parameters fixed at their experimental values at 4 K. The results using both spin-polarized GGA and GGA+U method with U$_{eff}$ = 4.0 eV are listed in Table 5, together with the neutron diffraction data. (Note that the internal structural parameters remain constant for U$_{eff}$ > 4.0 eV.) Both GGA and GGA+U results are in excellent agreement with the neutron diffraction data with the GGA+U method showing a slight improvement.

**Table 5.** Internal structural parameters optimized with spin polarized GGA and GGA+U methods with $U_{eff}$ = 4.0 eV, together with the neutron powder diffraction data at 4 K refined with Rietveld method. Only the free parameters are listed. The differences between the internal structural parameters optimized using first principles DFT based methods and determined by neutron powder diffraction are given within parentheses.

| Atom | Parameters | GGA | GGA+U ($U_{eff}$ = 4.0 eV) | Experiment |
|---|---|---|---|---|
| Rb | $x$ | 0.993 ( 0.002) | 0.990 ( 0.005) | 0.995 |
|  | $z$ | 0.372 ( 0.007) | 0.376 ( 0.003) | 0.379 |
| Fe(1) | $x$ | 0.808 (-0.007) | 0.804 (-0.003) | 0.801 |
|  | $z$ | 0.729 ( 0.001) | 0.730 ( 0.000) | 0.730 |
| F(1) | $x$ | 0.748 (-0.007) | 0.746 (-0.005) | 0.741 |
|  | $y$ | 0.073 (-0.009) | 0.066 (-0.002) | 0.064 |
|  | $z$ | 0.578 ( 0.000) | 0.576 ( 0.002) | 0.578 |
| F(2) | $x$ | 0.372 ( 0.007) | 0.380 (-0.001) | 0.379 |
|  | $y$ | 0.018 (-0.010) | 0.007 (0.001) | 0.008 |
|  | $z$ | 0.678 (-0.009) | 0.673 (-0.004) | 0.669 |
| F(3) | $x$ | 0.425 ( 0.008) | 0.422 ( 0.011) | 0.433 |
|  | $z$ | 0.451 ( 0.017) | 0.464 ( 0.004) | 0.468 |
| F(4) | $x$ | 0.576 (-0.007) | 0.574 (-0.005) | 0.569 |
|  | $z$ | 0.843 (-0.002) | 0.844 (-0.003) | 0.841 |

To elucidate the electronic structure of the system we have calculated and analyzed the partial density of states (PDOS) projected onto the 3d states of the two inequivalent iron ions, Fe(1) and F(2), and onto the F - 2$p$ states. These states are situated near the Fermi energy and are therefore expected to dominate the electronic and magnetic properties of the system. The PDOS are computed for the lowest energy collinear spin configuration – a partially frustrated antiferromagnetic configuration (see Discussion for details) – within GGA and GGA+U. As typically found for magnetic oxides/fluorides, GGA leads to a half metallic solution in which there is a gap at the Fermi level in the spin-up channel while the spin-down channel is metallic. The 3$d$ states of both Fe(1) and Fe(2), with significant admix with F-2$p$ states, show complete occupancy in the up spin channel (see Figure 5a). On the other hand in the spin-down channel the 3$d$ states of both Fe(1) and Fe(2) are partially filled. In order to show this change in occupancy of Fe-3$d$ states in the down spin channel more clearly, we show in Figures 5c and 5e the Fe-3$d$ PDOS and the integrated PDOS, respectively, plotted for a narrow energy window (-0.5 eV to 1.0 eV) around the Fermi level. As can been seen, Fe(1) $d$-states are slightly more occupied compared to Fe(2). This results in a slight charge disproportionation between Fe(1) and Fe(2), which is also reflected in the calculated magnetic moments of magnitude 3.83 $\mu_B$ and 3.97 $\mu_B$ at the Fe(1) and Fe(2) site respectively.

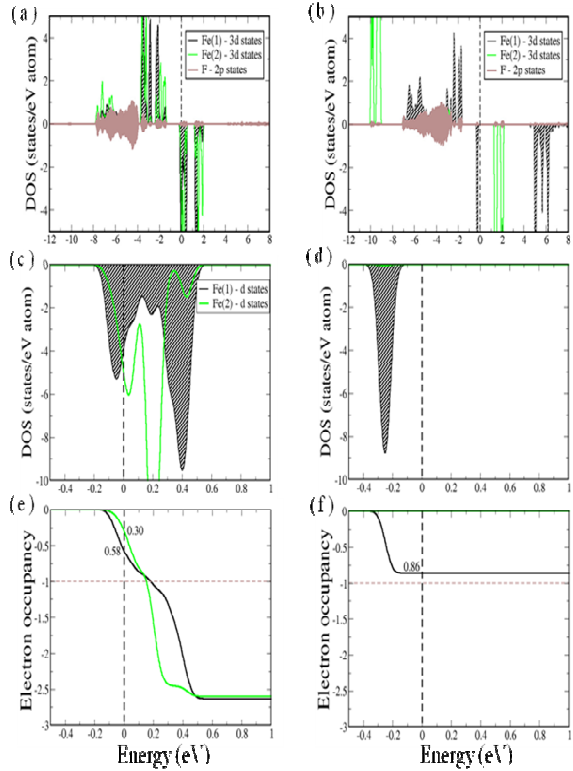

**Figure 5.** Upper panel and middle panel represent the partial density of states (PDOS) projected on to Fe(1)-$3d$ (shaded with black diagonal lines), Fe(2)-$3d$ (green solid line) and F-$2p$ (brown shaded region) states computed with GGA (a & c) and GGA+U (b & d) methods for $U_{eff}$ = 6.0eV, plotted for a large energy window of range -12.0 eV to 8.0 eV and for a small energy window around Fermi energy of range -0.5 eV to 1.0 eV, respectively. Third panel corresponds to the integrated DOS (IDOS) in the down-spin channel of Fe(1)-$3d$ PDOS (black solid line) and Fe(2)-$3d$ PDOS (green solid line). The black dashed vertical line corresponds to Fermi level and the brown horizontal line shows the occupancy = 1.0.

Within GGA+U ($U_{eff}$ = 6eV) the band gap of RbFe$_2$F$_6$ opens. The calculated value of $E_{gap}$ = 1.3eV is slightly underestimated from the measured optical band gap of 1.9 eV determined from UV-vis diffuse spectra. As shown in Figure 5b, the $3d$ states of both Fe(1) and Fe(2) also remain completely occupied in the up spin channel. The modification of the occupancy of the 3d states of Fe ions, however, is found to be significant in the down spin channel. (This remains true throughout the entire range of $U_{eff}$ = 1.0 to 7.0 eV as shown in the Supporting Information.) In Figure 5d and 5f we show the Fe-$3d$ PDOS and the integrated PDOS, respectively, plotted for a narrow energy window (-0.5 eV to 1.0 eV) around the Fermi level. With the increase of the on-site Coulomb interaction (U) a $3d$ state of Fe(1) is found to move towards the valence region and is completely occupied. The calculated occupancy of Fe(1)-$3d$ states in the down spin channel is found to be 0.86 (for all $U_{eff} \geq$ 4.0 eV), which is reduced from 1.00 attributable to hybridization with F-$2p$ states. This is consistent with a nominal oxidation state of Fe(1) being +2 and hence a $d^6$ electronic configuration. Whereas the completely empty spin-down channel for Fe(2) $3d$ states leads to the nominal valence of +3 and the $d^5$ electronic configuration. We therefore have an Fe$^{+2}$/Fe$^{+3}$ charge ordered ground state and an energy gap of the system that is between the $3d$ states of Fe(1) and Fe(2), which is in agreement with the nature of the optical gap obtained from UV-vis diffuse spectra. The charge separation between Fe(1) and Fe(2) is also reflected in the computed magnetic moments of 3.82 $\mu_B$ and 4.48 $\mu_B$ at Fe(1) and Fe(2) site respectively. Further strong evidence of a charge ordered ground state is found from a plot of the charge density within a small energy window below the Fermi energy (-0.5eV to 0.0 eV), within GGA and GGA+U as shown in Figures 6a and 6b respectively. Within GGA+U it can clearly be seen that an extra electron occupies an Fe(1) state in the spin-down channel compared with Fe(2).

**Discussion.**

**Synthesis.** Previously reported and related materials, e.g., NH$_4$Fe$_2$F$_6$, NH$_4$MnFeF$_6$, NH$_4$MnCrF$_6$, and RbMnFeF$_6$,[22, 50] were synthesized by using the binary metal fluorides mixed with the alkali metal fluoride of NH$_4$F solutions in a platinum tube. The tube was sealed, placed in an autoclave, and heated to temperatures above 350°C that resulted in pressures in excess of 2000 bar. We were able to synthesize RbFe$_2$F$_6$ through a low temperature and mild hydrothermal technique. In our method, the binary metal fluorides are combined with RbF and a diluted CF$_3$COOH aqueous solution. We have previously demonstrated that this method can be used to synthesize phase-pure and polycrystalline BaMF$_4$ (M = Mg, Mn, Co, Ni and Zn).[51]

**Variable Temperature Neutron Diffraction Data.** Figure 7 shows the lattice parameters from the variable temperature neutron diffraction data. All three lattice parameters show a modest contraction down to the magnetic ordering temperature, below which there is negative thermal expansion, particularly within the $ab$-plane. This is confirmed through the increase in volume, as shown in the inset to Figure 7b. No structural phase transition was found on cooling, and RbFe$_2$F$_6$ remained orthorhombic with *Pnma* symmetry to the lowest temperature measured. One central structural feature was the difference between thermal factors of the constituent elements. Figure 8 shows the isotropic thermal factor for all 6 atoms as a function of temperature. Although each atom has a typical temperature dependence, the absolute value of U for Rb is ≈ 5 times that of the lighter Fe, and even twice that of the F that is over four times lighter. The rattling effect of A site cations within the β-pyrochlores is well documented and results from a gross mismatch between the ionic radii of the cations and the available space. The magnetic structure was determined from powder neutron diffraction data collected at 4 K on the BT1 diffractometer at NIST. The new magnetic reflections that appeared below $T_N$ could all be indexed with a $k = (0\ 0\ 0)$ propagation vector. Representational analysis in the *Pnma* space group, calculated with the BASIREPS program, showed eight possible irreducible representations for the two iron atoms at Fe(1) (x 0.25 z) and Fe(2) (0.5 0 0.5). Each possibility was evaluated in the FULLPROF program [38] and a summary of the representations and comparison of the magnetic R-factors is given in Table 6.

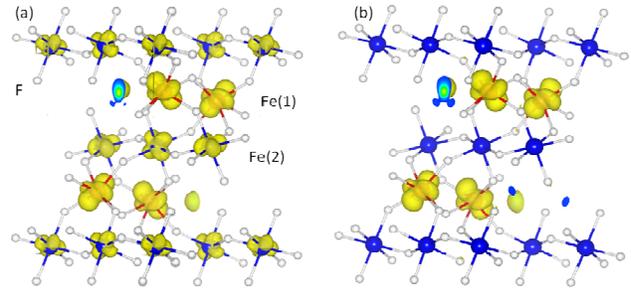

**Figure 6.** The electron charge density plotted for the energy range of -0.5 eV below the Fermi level, computed with GGA (a) and GGA+U (b) methods with $U_{eff}$ = 6.0 eV. The isovalue is set to 0.05 e/Å$^3$.

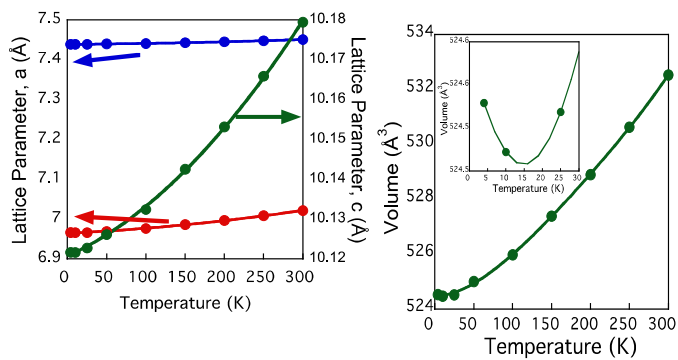

**Figure 7.** Panel a shows the lattice parameter as a function temperature; there is a contraction in all three directions upon cooling, until the magnetic ordering temperature when RbFe$_2$F$_6$ shows negative thermal expansion in a, b and c best visualized by panel b the overall increase in volume.

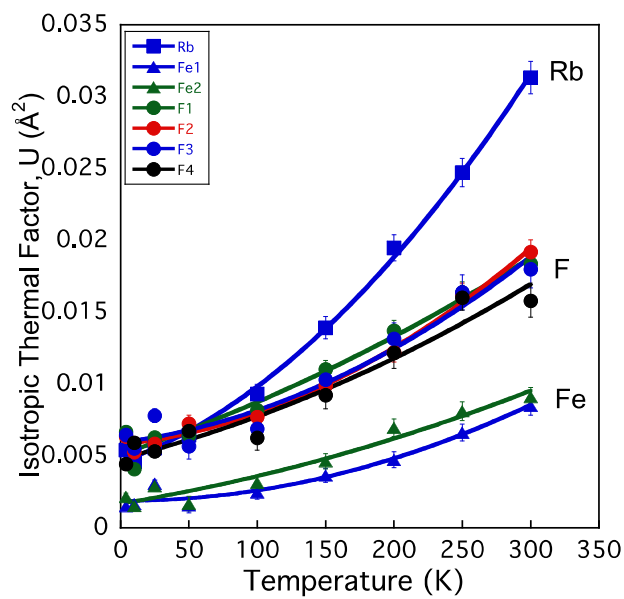

**Figure 8.** Isotropic thermal factors of RbFe$_2$F$_6$ as a function of temperature as obtained from Rietveld refinement of powder neutron diffraction data showing the extraordinary large thermal factor of Rb as a result of the mismatch between Rb ionic size and the open cage site that it occupied.

**Table 6.** Irreducible representations of Fe(1) at (x 0.5 z) and Fe(2) at (0.5 0 0.5) in the Pnma space group and a (0 0 0) propagation vector are shown. A comparative magnetic R-factor before full refinement is given in the final column and represents the experimental determined magnetic structure.

| Irrep. Rep. | Atom Position | Basis Vector for Fe (0.80 0.25 0.73) | Basis Vector for Fe (0.5 0 0.5) | $R_{mag}$ |
|---|---|---|---|---|
| $\Gamma_1$ | (x, y, z) (x+½, y, z+½) (x, y+½, z) (x+½, y+½, z+½) | (0 1 0) (0 -1 0) (0 1 0) (0 -1 0) | (1 0 0)(0 1 0)(0 0 1) (-1 0 0)(0 -1 0)(0 0 1) (-1 0 0)(0 1 0)(0 0 -1) (1 0 0)(0 -1 0)(0 0 1) | 8.82 |
| $\Gamma_2$ | (x, y, z) (x+½, y, z+½) (x, y+½, z) (x+½, y+½, z+½) | (1 0 0)(0 0 1) (-1 0 0)(0 0 1) (-1 0 0)(0 0 -1) (1 0 0)(0 0 -1) | | 88.1 |
| $\Gamma_3$ | (x, y, z) (x+½, y, z+½) (x, y+½, z) (x+½, y+½, z+½) | (1 0 0)(0 0 1) (-1 0 0)(0 0 1) (1 0 0)(0 0 1) (-1 0 0)(0 0 1) | (1 0 0)(0 1 0)(0 0 1) (-1 0 0)(0 -1 0)(0 0 1) (1 0 0)(0 -1 0)(0 0 1) (-1 0 0)(0 1 0)(0 0 1) | 13.1 |
| $\Gamma_4$ | (x, y, z) (x+½, y, z+½) (x, y+½, z) (x+½, y+½, z+½) | (0 1 0) (0 -1 0) (0 -1 0) (0 1 0) | | 104 |
| $\Gamma_5$ | (x, y, z) (x+½, y, z+½) (x, y+½, z) (x+½, y+½, z+½) | (0 1 0) (0 1 0) (0 1 0) (0 1 0) | (1 0 0)(0 1 0)(0 0 1) (1 0 0)(0 1 0)(0 0 -1) (-1 0 0)(0 1 0)(0 0 -1) (-1 0 0)(0 1 0)(0 0 1) | 19.3 |
| $\Gamma_6$ | (x, y, z) (x+½, y, z+½) (x y+½, z) (x+½, y+½, z+½) | (1 0 0)(0 0 1) (1 0 0)(0 0 -1) (-1 0 0)(0 0 -1) (-1 0 0)(0 0 1) | | 88 |
| $\Gamma_7$ | (x, y, z) (x+½, y, z+½) (x, y+½, z) (x+½, y+½, z+½) | (1 0 0)(0 0 1) (1 0 0)(0 0 -1) (1 0 0)(0 0 1) (1 0 0)(0 0 -1) | (1 0 0)(0 1 0)(0 0 1) (1 0 0)(0 1 0)(0 0 -1) (1 0 0)(0 -1 0)(0 0 1) (1 0 0)(0 -1 0)(0 0 -1) | 20.8 |
| $\Gamma_8$ | (x, y, z) (x+½, y, z+½) (x, y+½, z) (x+½, y+½, z+½) | (0 1 0) (0 1 0) (0 -1 0) (0 -1 0) | | 81.2 |

$\Gamma_1$ was clearly identifiable as giving the best fit to the experimental data. Within this representation, the moments of Fe at (x 0.25 z) are confined to the *b* axis, whereas the second Fe position at (0.5 0 0.5) is allowed to have components along either the *a*, *b* or *c* axis. However, refinement of the moment on the second Fe position only gave contributions along the *a* axis, and no improvement to the fit of the powder neutron diffraction data could be achieved with additional components along *b* or *c*. These latter components were omitted from the final refinements. The atomic positions of the final refinements are given in Table 7 and the observed, calculated and difference plots are shown in Figure 9. The magnetic structure is shown in Figure 10. Fe(1) with a magnetic moment of 3.99(5) $\mu_B$ is confined to the *b* axis, but can be described as forming antiferromagnetic chains along the *a*-axis. These chains are orthogonal to those on Fe(2). The second Fe moment at (0.5 0 0.5) has a slightly larger magnitude of 4.29(5) $\mu_B$, consistent with its $Fe^{3+}$ oxidation state, and resides along the *a*-axis forming antiferromagnetic chains parallel to the *b*-axis. The refined moments are all smaller than the theoretical spin-only contribution to the magnetic moment, but this is consistent with the observed diffuse scattering that is present even at 4 K, resulting from the magnetic frustration, suggesting that not all of the moments become long-range ordered. The refined moments are, however, consistent with the GGA+U calculated moments, $Fe(1)_{calc(exp)}$ = 3.82$\mu_B$ (3.99(5)$\mu_B$) and $Fe(2)_{calc(exp)}$ = 4.48$\mu_B$ (4.29(5)$\mu_B$).

| Atom | x | y | z | B | Moment |
|---|---|---|---|---|---|
| Rb | 0.9942 (3) | 0.25 (-) | 0.3798 (2) | 0.43 (3) | |
| Fe1 | 0.8014 (2) | 0.25 (-) | 0.7302 (1) | 0.20 (2) | 3.99 (5) |
| Fe2 | 0.5 | 0.0 | 0.5 | 0.17 (2) | 4.29 (5) |
| F1 | 0.7413 (2) | 0.0644 (2) | 0.5784 (2) | 0.53 (3) | |
| F2 | 0.3791 (2) | 0.0077 (3) | 0.6698 (1) | 0.50 (3) | |
| F3 | 0.4322 (3) | 0.25 | 0.4672 (2) | 0.51 (4) | |
| F4 | 0.5682 (3) | 0.25 | 0.8407 (2) | 0.35 (3) | |

**Table 7.** Refined atomic positions as obtained from Rietveld refinement of the powder neutron diffraction at 4 K in the Pnma space group with cell parameters of a = 6.96630 (5) Å, b = 7.43903 (5) and c = 10.12164 (7) Å. Final wRp = 4.34 %, RBragg = 2.17 % and a magnetic R-factor of 3.35%.

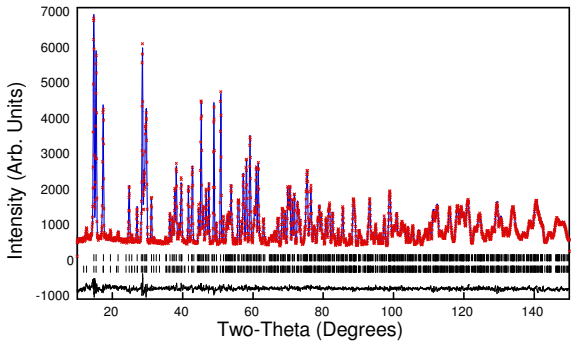

**Figure 9.** Observed (red), calculated (blue) and difference (black) data obtained from Rietveld refinements of neutron diffraction data of $RbFe_2F_6$ at 4 K. The upper tickmarks represent those associated with the nuclear structure while those below correspond to the magnetic structure refinement.

**Magnetism.** Considerable additional information can be gathered from the $\chi^{-1}$ vs. *T* plot shown in Figure 4b. The data are seen to adhere quite well to the Curie-Weiss (C-W) form ($\chi = C/T-\theta$), where *C* and $\theta$ are constants) for *T* > 100 K or so, yielding a Weiss temperature of -272 K. Fits with an additional temperature-independent paramagnetic susceptibility describe the data even better, yielding an effective number of Bohr magnetons of 7.9 $\mu_B$ / f.u. Note that measurement of $\chi(T)$ in magnetic fields in the range 10–$10^4$ Oe yielded magnetic moment and $\theta$ values that varied by only 10-20 %, consistent with the fact that the *M(H)* curves are quite linear at all *T* (see inset to Figure 4b). The extracted values are similarly robust with respect to the exact temperature range used for the fitting to the C-W form. The theoretical spin only value is 7.7 $\mu_B$ / f.u. ($Fe^{2+}$ = 4.9 $\mu_B$, $Fe^{3+}$ = 5.9 $\mu_B$), in good agreement with the data.. Importantly, the large negative Weiss temperature indicates relatively strong AF interactions between the Fe moments. In fact, comparison to the actual AF ordering temperature of 16 K indicates significant magnetic frustration in this compound, with a frustration ratio ($\theta$ / $T_N$) of 17.[52]

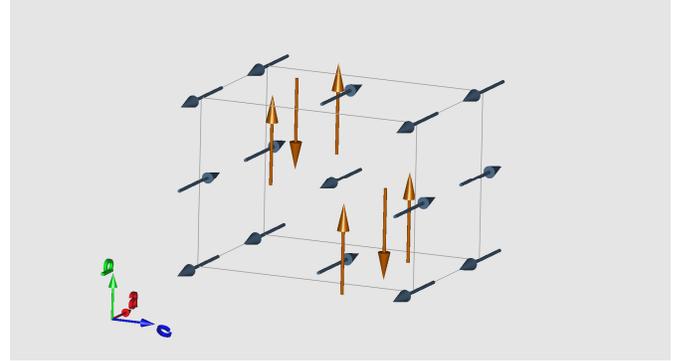

**Figure 10.** Magnetic structure of $RbFe_2F_6$ at 4 K as determined from Rietveld refinement of powder neutron diffraction data. The magnetic structure has a (0 0 0) propagation vector with the Fe(1) moment (gold) aligned along the *b* axis forming antiferromagnetic chains down *a*, whereas the Fe(2) moments (blue) align along the *a* axis and form antiferromagnetic chains along *b*.

**Theoretical Calculations.** We elucidate the magnetic properties of $RbFe_2F_6$ by expanding the total energy of the system using the Heisenberg model,

$$E = E_0 - \sum_{ij} J_{ij} S_i . S_j \qquad (1)$$

where $E_0$ is the total energy of the system in the orthorhombic symmetry without spin-spin interaction. Here $S_i$ is the spin of the $i^{th}$ Fe ion and $J_{ij}$ denotes the exchange integral between $i^{th}$ and $j^{th}$ Fe ions. A negative (positive) value of *J* corresponds to antiferromagnetic (ferromagnetic) nature of the coupling. The magnetic exchange integrals *J* can be found by fitting the total energy equations for eight collinear spin configurations assuming *S* = 2 and *S* = 5/2 for Fe(1) and Fe(2) ions respectively. We have considered all the nearest neighbor (NN) exchange integrals around each Fe ion. The exchange integral pathways that we considered in the expansion of the total energy are given in Figure 11a. It is important to note that these paths construct an Fe$_4$ tetrahedral unit, which is the basic building block of the magnetic network of the system. Therefore the system is expected to be completely geometrically frustrated when all these NN interactions, $J_1$ to $J_4$, are antiferromagnetic in nature and exactly equal in magnitude to each other. The nature, as well as the magnitudes of these exchange integrals therefore plays a crucial role to drive the stability of the magnetic ground state of the system.

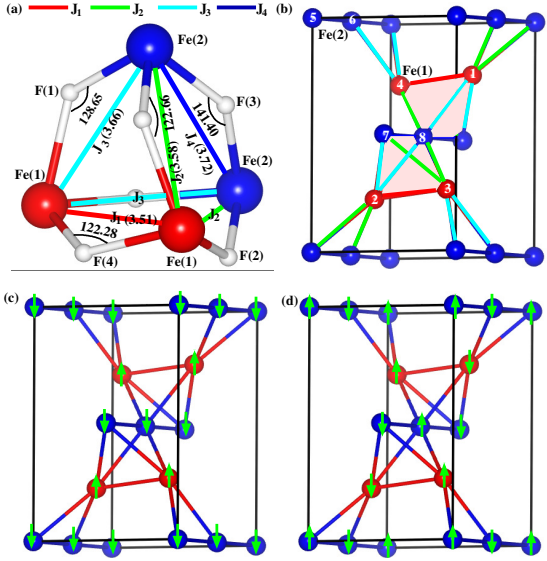

**Figure 11.** In the upper panel: (a) represents the Fe$_4$ tetrahedron and (b) represents the magnetic unit cell. The Fe-Fe distances are in Å given inside the parentheses. The pink shaded regions show the Fe$_4$ tetrahedron unit. Eight Fe ions in the orthorhombic unit cell have been numbered. In the lower panel: (c) and (d) show SC-2 and SC-4 respectively.

The total energy results for each of the eight spin configurations are described in Table 8 where we find in both GGA and GGA+U that the ground state is antiferromagnetic (AFM). The minimum collinear AFM spin configuration (SC) energy, however, is found to switch between two AFM configurations, SC-2 to SC-4, when we include finite correlation effects (we calculate the ground state to be SC-4 for all $U_{eff}$ > 1eV). The main difference between these two states is the coupling within a chain of spins. For SC-2, the spins are coupled ferromagnetically along both the Fe(1) and the Fe(2) chains, with the coupling between chains being antiferromagnetic (see Figure 11c). In contrast to this, in SC-4, spins are antiferromagnetically coupled along both the Fe(1) and the Fe(2) chains, while the chains remain antiferromagnetically coupled to each other (see Figure 11d). SC-4 is close to the experimentally determined non-collinear spin structure (see Figure 10).

**Table 8.** Eight spin configurations of the Fe ions within the orthorhombic unit cell used to determine the lowest energy collinear spin configuration and the magnetic exchange integrals, $J_1$ to $J_4$ (as described in Figures 11a and b), where $E1 = -J_1S_1S_1$, $E2 = -J_2S_1S_2$, $E3 = -J_3S_1S_2$ and $E4 = -J_4S_2S_2$. $S_1$ and $S_2$ represent the spin of Fe(1) and F(2) respectively. + and – sign correspond to spin up and down respectively. Numbering of the magnetic ions follow Figure 11b. In the last two columns the calculated total energies of the collinear spin configurations obtained using GGA and GGA+U method with $U_{eff}$ = 6.0 eV are given respectively, with respect to the spin configuration 1, where all spins are pointing along same direction.

| Confg. | Magnetic ions | | | | | | | | Total energy equations | Calculated total energy (meV) | |
|---|---|---|---|---|---|---|---|---|---|---|---|
| | 1 | 2 | 3 | 4 | 5 | 6 | 7 | 8 | | GGA | GGA+U 6.0eV |
| 1 | + | + | + | + | + | + | + | + | $E0 + 4E1 + 8E2 + 8E3 + 4E4$ | 0 | 0 |
| 2 | + | + | + | + | – | – | – | – | $E0 + 4E1 - 8E2 - 8E3 + 4E4$ | -437 | -172 |
| 3 | + | – | – | + | + | + | – | – | $E0 + 4E1 + 4E4$ | -218 | -88 |
| 4 | – | – | + | + | + | – | – | + | $E0 - 4E1 - 4E4$ | -422 | -200 |
| 5 | + | – | + | – | + | + | – | – | $E0 - 4E1 + 4E4$ | -239 | -114 |
| 6 | + | + | + | + | – | + | + | + | $E0 + 4E1 + 4E2 + 4E3$ | -171 | -87 |
| 7 | + | – | + | + | – | + | + | + | $E0 + 4E2$ | -199 | -130 |
| 8 | + | – | + | + | – | – | + | + | $E0 + 4E2 - 4E3 + 4E4$ | -276 | -118 |

From these total energy calculations the exchange integrals were extracted. All were found to be antiferromagnetic in nature within both GGA and GGA+U methods as listed in Table 9. For GGA the third coupling, $J_3$, which is the coupling between two NN Fe$^{+2}$ (Fe(1)) and Fe$^{+3}$ (Fe(2)) ions through F(1) ion, is found to be the strongest. The second strongest coupling is $J_4$, which accounts for the coupling between two NN Fe$^{+3}$ ions mediated via F (3) ion, is 0.85 of $J_3$. However within GGA+U, this $J_4/J_3$ ratio gets enhanced with the increase of U and becomes more than 1.0 for $U_{eff}$ > 2.0 eV. The coupling between two NN Fe$^{+2}$ ions, $J_1$, and the coupling between two NN Fe$^{+2}$ and Fe$^{+3}$ mediated via F(2) ion, $J_2$, remain smaller than $J_3$ and $J_4$ (for the whole range of U). These results indicate that the magnetic frustration is partially lifted by the orthorhombic distortion, but not completely, as $J_1$ and $J_2$ are not negligible compared to $J_3$ and $J_4$. This residual frustration is therefore a hindrance to achieve a collinear AFM ordering for the system and likely leads to a non-collinear AFM state governed by the strongest exchange integrals.

**Table 9.** The obtained magnetic exchange integrals in meV and the mean field estimation of Curie-Weiss temperature ($\Theta_{CW}$) in K are listed.

| $J_i$ | GGA | GGA+U 6.0 eV |
|---|---|---|
| $J_1$ | -1.8 | -0.9 |
| $J_2$ | -2.7 | -0.7 |
| $J_3$ | -4.0 | -1.5 |
| $J_4$ | -3.4 | -1.7 |
| $\Theta_{CW}$ | -693 | -281 |

Finally we calculated the Curie-Wiess temperature, $\Theta_{CW}$, within mean field theory given by,

$$\Theta_{CW} = \frac{1}{3k_B}(2S_1(S_1+1)J_1 + 2S_{12}(S_{12}+1)(J_2+J_3) + 2S_2(S_2+1)J_4) \quad (2)$$

where $S_{12}$ is the average of $S_1$ and $S_2$. Results are listed in Table 9 where it can be seen that the GGA+U result, $\Theta_{CW}$ = -281K, agrees well with the experimental value of -272K, extracted from magnetic susceptibility data at high temperatures (Note that this is why we choose to present the U = 6eV results.)

**Conclusion.**

We have synthesized and characterized a new charge-ordered magnetically frustrated mixed-metal fluoride, $RbFe^{2+}Fe^{3+}F_6$, that exhibits a pyrocholore-related structure. An anti-ferromagnetic ordering temperature of 16K was observed, however no structural transition was observed in the variable temperature neutron diffraction data. The theoretical calculations for the Weiss constant resulted in a value of -281K that agrees well with the experimental value of -272K. Additional theoretical calculations, neutron diffraction, magnetic, and Mossbauer measurements on $RbFe_2F_6$ are in progress and will be reported in the near future.

**ACKNOWLEDGMENT.** Work at UH and Cornell supported by the U.S. Department of Energy, Basic Energy Sciences, Division of Materials Sciences and Engineering under Award # DE-SC0005032. Work at UMN supported by DOE Award #DE-FG02-06ER46275 (magnetic measurements).

**Notes and references**

[a] *Department of Chemistry, University of Houston, 136 Fleming Building, Houston, TX 77204-5003, USA: Fax:+011 713-743-0796: Tel: +011 713-743-3278: E-mail: psh@uh.edu*
[b] *Department of Materials Science and Engineering, University of Maryland, College Park, MD, 20742-2115, USA and NIST Center for Neutron Research, National Institute of Standards and Technology, 100 Bureau Drive, Gaithersburg, MD 20899-6103, USA:. Fax:+011 301-975-4297: Tel: +011301-975-4297: E-mail: mark.green@nist.gov*
[c]*Department of Chemical Engineering and Materials Science, University of Minnesota, 151 Amundson Hall, 421 Washington Ave. SE, Minneapolis, MN 55455, USA: Fax: +011 612-624-4578 Tel: +011 612-625-4018: E-mail: Leighton@umn.edu*
[d]*School of Applied and Engineering Physics, 271 Clark Hall, Cornell University, Ithaca, NY 14853, USA: Fax:+011 607-255-7658 Tel: +011 607-255-6498: E-mail: fennie@cornell.edu*

† Electronic Supplementary Information (ESI) available: X-ray crystallographic files in CIF format, experimental and calculated powder X-ray diffraction patterns, Infrared and UV-vis spectra, thermogravimetric and differential thermal analysis diagrams, and additional theoretical calculations. See DOI: 10.1039/b000000x/

1   P. Hagenmuller, in *Book Inorganic Solid Fluorides - Chemistry and Physics*, ed., ed. by Editor, Academic Press, City, **1985**.
2   M. Eibschutz, H. J. Guggenheim, *Solid State Commun.* 1968, **6**, 737-739.
3   M. Eibschutz, H. J. Guggenheim, S. H. Wemple, I. Camlibel, M. DiDomenico, *Phys. Lett.* 1969, **29A**, 409-410.
4   E. T. Keve, S. C. Abrahams, J. L. Bernstein, *J. Chem. Phys.* 1969, **51**, 4928-4936.
5   E. T. Keve, S. C. Abrahams, J. L. Bernstein, *J. Chem. Phys.* 1970, **53**, 3279-3287.
6   J. Ravez, S. Arquis, J. Grannec, *J. Appl. Phys.* 1987, **62**, 4299-4301.
7   R. Cosier, A. Wise, A. Tressaud, J. Grannec, R. Olazcuaga, J. Portier, *C. R. Hebd. Seances Acad. Sci. Ser. C* 1970, **271**, 142.
8   G. Ferey, R. De Pape, M. Poulain, D. Grandjean, A. Hardy, *Acta Cryst.* 1977, **B33**, 209.
9   J. Ravez, S. C. Abrahams, R. De Pape, *J. Appl. Phys.* 1989, **65**, 3987.
10  J. Ravez, *J. Phys. III* 1997, **7**, 1129.
11  S. C. Abrahams, J. Ravez, A. Simon, J. P. Chaminade, *J. Appl. Phys.* 1981, **57**, 4740-4743.
12  J. G. Bergman, G. R. Crane, H. J. Guggenheim, *J. Appl. Phys.* 1975, **46**, 4645-4646.
13  K. Shimamura, E. G. Villora, H. Zeng, M. Nakamura, S. Takekawa, K. Kitamura, *Appl. Phys. Lett.* 2006, **89**, 232911-1 - 232911-3.
14  R. Blinc, G. Tavcar, B. Zemva, D. Hanzel, P. Cevc, C. Filipic, A. Levstik, Z. Jaglicic, Z. Trontelj, N. Dalal, V. Ramachandran, S. Nellutla, J. F. Scott, *J. Appl. Phys.* 2008, **103**, 074114-1 - 074114-4.
15  R. Blinc, G. Tavcar, B. Zemva, E. Goreshnik, D. Hanzel, P. Cevc, A. Potocnik, V. Laguta, Z. Trontelj, Z. Jaglicic, J. F. Scott, *J. Appl. Phys.* 2009, **106**, 023924-1 - 023924-4.
16  R. Blinc, P. Cevc, A. Potocnik, B. Zemva, E. Goreshnik, D. Hanzel, A. Gregorovic, Z. Trontelj, Z. Jaglicic, V. Laguta, M. Perovic, N. Dalal, J. F. Scott, *J. Appl. Phys.* 2010, **107**, 043511-1 - 043511-5.
17  J. F. Scott, R. Blinc, *J. Phys.: Condens. Matter* 2011, **23**, 1 - 17.
18  J. F. Scott, R. Blinc, *J. Phys. Condens. Matter* 2011, **23**, 299401.
19  D. Babel, G. Pausewang, W. Viebahn, *Z. Naturforsch.* 1967, **22b**, 1219-1220.
20  D. Babel, *Z. Anorg. Allg. Chem.* 1972, **387**, 161-178.
21  N. N. Greenwood, A. T. Howe, F. Menil, *J. Chem. Soc. A* 1971, 2218-2224.
22  G. Ferey, M. Leblanc, R. De Pape, *J. Solid State Chem.* 1981, **40**, 1-7.
23  A. Tressaud, F. Menil, R. Georges, J. Portier, P. Hagenmuller, *Mat. Res. Bull.* 1972, **7**, 1339-1346.
24  N. N. Greenwood, F. Menil, A. Tressaud, *J. Solid State Chem.* 1972, **5**, 402-409.
25  A.-M. Hardy, A. Hardy, G. Ferey, *Acta Cryst.* 1973, **B29**, 1654-1658.
26  R. Masse, J. Aleonard, M. T. Averbuch-Pouchot, *J. Solid State Chem.* 1984, **53**, 136.
27  J. Gaile, W. Rudorff, W. Viebahn, *Z. Anorg. Allg. Chem.* 1977, **430**, 161.
28  S. Roth, W. Kurtz, *Physica* 1977, **86-88B**, 715.

29   J. Villain, *Z. Phys. B* 1979, **33**, 31.
30   G. Ferey, M. Leblanc, R. de Pape, J. Pannetier, *Solid State Commun.* 1985, **53**, 559-563.
31   G. Ferey, M. Leblanc, R. De Pape, *J. Cryst. Growth* 1975, **29**, 209-211.
32   E. Herdtweck, *Z. Anorg. Allg. Chem.* 1983, **501**, 131.
33   M. Leblanc, G. Ferey, Y. Calage, R. De Pape, *J. Solid State Chem.* 1984, **53**, 360.
34   SAINT, Program for Area Detector Absorption Correction, Siemens Analytical X-ray Systems, Inc., Madison, WI, **1995**.
35   G. M. Sheldrick, *SHELXS-97 - A program for automatic solution of crystal structures. University of Goettingen: Goettingen, Germany, 1997.*
36   G. M. Sheldrick, *SHELXL-97 - A program for Crystal Structure Refinement; University of Goettingen: Goettingen, Germany, 1997.*
37   L. J. Farrugia, *J. Appl. Crystallogr.* 1999, **32**, 837-838.
38   J. Rodriguez-Carvajal, *Phys. B* 1993, **192**, 55.
39   P. M. Kubelka, F. Z. Munk, *Tech. Phys.* 1931, **12**, 593.
40   J. Tauc, *Mater. Res. Bull.* 1970, **5**, 721-9.
41   J. P. Perdew, K. Burke, M. Ernzerhof, *Phys. Rev. Lett.* 1996, **77**, 3865.
42   P. E. Blöochl, *Phys. Rev. B* 1994, **50**, 17953.
43   G. Kresse, J. Hafner, *Phys. Rev. B* 1993, **47**, 558.
44   G. Kresse, J. Furthmuller, *Phys. Rev. B* 1996, **54**, 1116.
45   V. I. Anisimov, F. Aryasetiawan, A. I. Lichtenstein, *J. Phys. Condens. Matter* 1997, **9**, 767.
46   S. L. Dudarev, G. A. Botton, S. Y. Savrasov, C. J. Humphreys, A. P. Sutton, *Phys. Rev. B* 1998, **57**, 1505.
47   I. D. Brown, D. Altermatt, *Acta Crystallogr., Sect. B* 1985, **B41**, 244-7.
48   N. E. Brese, M. O'Keeffe, *Acta Crystallogr., Sect. B* 1991, **B47**, 192-7.
49   K. Nakamoto, *Infrared and Raman Spectra of Inorganic and Coordination Compounds Part A: Theory and Applications in Inorganic Chemistry*. 5th ed., Editor, John Wiley & Sons, Inc., New York, 1997.
50   M. Leblanc, G. Ferey, Y. Calage, R. de Pape, *J. Solid State Chem.* 1983, **47**, 24-29.
51   S. W. Kim, H. Y. Chang, P. S. Halasyamani, *J. Am. Chem. Soc.* 2010, **132**, 17684-17685.
52   A. P. Ramirez, *Annu. Rev. Mater. Sci.* 1994, **24**, 453.